\documentclass[aps,prb,preprint,groupedaddress,citeautoscript,nopacs]{revtex4}

\usepackage{graphicx}
\usepackage{color}
\usepackage{amsfonts}
\setcitestyle{super}

\begin{document}

\title{Classification of stable three-dimensional Dirac semimetals with nontrivial topology}

\author{Bohm-Jung Yang$^{1}$ and Naoto Nagaosa$^{1,2}$}

\affiliation{$^1$ RIKEN Center for Emergent Matter Science (CEMS), Wako, Saitama 351-0198, Japan}

\affiliation{$^2$ Department of Applied Physics, University of Tokyo, Tokyo 113-8656, Japan}

\date{\today}

\begin{abstract}
A three-dimensional (3D) Dirac semimetal is the 3D analog of graphene
whose bulk band shows a linear dispersion relation in the 3D momentum space.
Since each Dirac point with four-fold degeneracy carries a zero Chern number, a Dirac semimetal can be stable
only in the presence of certain crystalline symmetries.
In this work, we propose a general framework to classify stable 3D Dirac semimetals.
Based on symmetry analysis, we show that various types of stable 3D Dirac semimetals
exist in systems having the time-reversal, inversion, and uniaxial rotational symmetries.
There are two distinct classes of stable 3D Dirac semimetals.
In the first class,
a pair of 3D Dirac points locate on the rotation axis, away from its center.
Moreover, the 3D Dirac semimetals in this class have nontrivial topological
properties characterized by 2D topological invariants,
such as the $Z_{2}$ invariant or the mirror Chern number.
These 2D topological indices give rise to stable 2D surface Dirac cones, which can be deformed
to Fermi arcs when the surface states couple to the bulk states on the Fermi level.
On the other hand, the second class of Dirac SM phases possess a 3D Dirac point
at a time-reversal invariant momentum (TRIM) on the rotation axis
and do not have surface states in general.
\end{abstract}


\maketitle

A Dirac semimetal (SM) indicates a phase
whose low energy excitations can be described by the pseudorelativistic Dirac fermions
with the linear energy dispersion.
Before the discovery of three dimensional (3D) topological insulators,~\cite{TI_Kane,TI_Zhang}
graphene~\cite{Graphene} has been considered as the unique system where the intriguing properties of the two dimensional (2D) Dirac fermions can be observed.
However, the recent progress in the field of topological insulators has shown that
stable 2D Dirac fermions exist ubiquitously on the surface of 3D topological insulators.~\cite{TI_Kane,TI_Zhang}
Moreover, through the careful studies on the topological phase transition
between a 3D topological insulator and a normal insulator,~\cite{Murakami0,Murakami1,Murakami2,TPT_Hasan,TPT_Ando}
it is demonstrated that even the 3D Dirac fermions with the linear dispersion in all three momentum directions can be observed
in the same material if we can reach the quantum critical point.
Since the 3D Dirac point with four-fold degeneracy does not carry a topological number,
the degeneracy at the gap-closing point can be easily lifted by small external perturbations,
hence the 3D Dirac fermions can be observed only at the single quantum critical point.
However, the approach to the quantum critical point
requires the intricate fine-tuning of the alloy chemical compositions,~\cite{TPT_Hasan,TPT_Ando}
which limits the accessibility to the fascinating physics of 3D Dirac fermions in experiments.

The breakthrough in the search for stable 3D Dirac semimetals
is achieved in the recent series of studies on Na$_{3}$Bi~\cite{Na3Bi_Shen, Na3Bi_Hasan}
and Cd$_{3}$As$_{2}$~\cite{Cd3As2_Cava, Cd3As2_Hasan, Cd3As2_Yazdani} compounds
where a pair of 3D bulk Dirac points stably exist on the $k_{z}$ axis.
The stability of the 3D Dirac points in these materials stems
from the fact that the system has additional crystalline symmetries
other than the time-reversal symmetry (TRS) and inversion symmetry (IS).
For instance, Young et al.~\cite{theory_Young,theory_Young2} have proposed that particular space groups allow
3D Dirac points as symmetry protected degeneracies. Also
Wang et al. have shown the symmetry protection of the 3D Dirac points through
the detailed symmetry analysis of Na$_{3}$Bi~\cite{LDA_Wang1} and Cd$_{3}$As$_{2}$~\cite{LDA_Wang2}.

In the present paper, we propose a general framework to classify stable 3D Dirac SMs
in systems with TRS, IS and uniaxial rotation symmetry, which are the most common symmetries of crystalline solids.
Through the careful examination of the condition for the accidental band crossing (ABC),
we have accomplished the complete classification of stable 3D Dirac SMs,
and uncovered that there is a class of 3D Dirac SMs which have nontrivial topological properties.
In particular, we have clarified the fundamental relationship between
the crystalline symmetries and the topological property of 3D Dirac SMs,
and demonstrated that the 3D {\it topological} Dirac SM generally mediates the topological quantum phase transition
between a 3D {\it weak} topological insulator and
a normal insulator. In fact,
the 3D topological Dirac SM itself is a parent state of 3D topological insulators,
which turns into either a 3D strong topological insulator~\cite{LDA_Wang1,LDA_Wang2}  or a 3D topological crystalline insulator
when the 3D Dirac point acquires a mass gap due to symmetry breaking.

\subsection*{Results}
{\bf Basic principles to create 3D Dirac SMs.}
Our strategy to synthesize a 3D Dirac SM is as follows. Let us consider
a system having both the TRS and IS.
In general, the TRS requires $E_{n,\uparrow}(\textbf{k})=E_{n,\downarrow}(-\textbf{k})$
where $E_{n,\sigma}(\textbf{k})$ indicates the energy eigenvalue of the $n$-th band with the spin $\sigma=\uparrow,\downarrow$
at the momentum $\textbf{k}$.
On the other hand, the IS requires $E_{n,\sigma}(\textbf{k})=E_{n,\sigma}(-\textbf{k})$.
Therefore under the combined operation of TRS and IS, $E_{n,\uparrow}(\textbf{k})=E_{n,\downarrow}(\textbf{k})$,
hence the energy band is doubly degenerate locally at each $\textbf{k}$.
Under this condition, whenever an ABC occurs between the valence and conduction bands,
a 3D Dirac point (DP) with four-fold degeneracy can be generated.
According to Murakami et al.,~\cite{Murakami0,Murakami1,Murakami2}
such an ABC can be achieved only under certain limited conditions
because of the strong repulsion between degenerate bands.
Namely,
only when the valence and conduction bands have the opposite parities, an ABC can occur
at a time-reversal invariant momentum (TRIM) by tuning an external parameter $m$.
In this case, a 3D DP appears at the quantum critical point ($m=m_{c}$)
between a normal insulator and a $Z_{2}$ topological insulator. (See Figure 1a.)
Since a band gap opens immediately once $m\neq m_{c}$, the DP is unstable.
However, in many crystals, the rotational symmetry as well as TRS and IS present
ubiquitously and constrain the physical properties of materials.
Surprisingly, as we will describe in detail below,
the additional uniaxial rotational symmetry strongly modifies the condition for ABC, which allows the 3D Dirac SM to emerge as {\it a stable phase}.
This is possible because when the valence and conduction bands have different rotation eigenvalues,
the level repulsion between them can be significantly relaxed as pointed out by Wang et al.,~\cite{LDA_Wang1,LDA_Wang2}
which eventually leads to the emergence of a stable 3D Dirac SM phase in the wide range of the parameter space. (See Figure 1b and c.)


To describe an ABC of two bands, each of which is doubly degenerate due to
the simultaneous presence of TRS and IS, a $4\times 4$ matrix Hamiltonian
can be used as a minimal Hamiltonian, which in general
has the following form,
\begin{displaymath}\label{eqn:minimalH}
H(\textbf{k})=
\sum_{i,j=0}^{3}a_{ij}(\textbf{k})\sigma_{i}\tau_{j}
=
\left( \begin{array}{cc}
h_{\uparrow\uparrow}(\textbf{k}) & h_{\uparrow\downarrow}(\textbf{k}) \\
h_{\downarrow\uparrow}(\textbf{k}) & h_{\downarrow\downarrow}(\textbf{k})
\end{array} \right),
\end{displaymath}
where the Pauli matrix $\sigma_{1,2,3}$ ($\tau_{1,2,3}$)
indicates the spin (orbital) degrees of freedom and $\sigma_{0}$ and $\tau_{0}$
are the $2\times 2$ identity matrices.
$h_{\sigma\sigma'}$ ($\sigma,\sigma'$=$\uparrow,\downarrow$)
indicates a $2\times2$ matrix which can be spanned by
$\tau_{0,1,2,3}$ and $a_{ij}(\textbf{k})$ are real functions of $\textbf{k}$.
The invariance of the system under the $C_{n}$ rotation gives
$C_{n}H(\textbf{k})C_{n}^{-1}=H(R_{n}\textbf{k})$
where $R_{n}$ is the $3\times3$ rotation matrix defining
the $2\pi/n$ rotation in the 3D space.~\cite{Fang}
Without loss of generality, we can choose the $k_{z}$ axis
as the axis of the $C_{n}$ rotation.
Then along the $k_{z}$ axis
on which $R_{n}\textbf{k}=\textbf{k}$ is satisfied,
$[C_{n},H(\textbf{k})]=0$. Therefore
we can choose a basis in which both $H(k_{z})|_{k_{x}=k_{y}=0}$ and $C_{n}$
are diagonal, hence
all bands
on the $k_{z}$ axis can be labeled by
the corresponding eigenvalues of $C_{n}$.
In such a basis,
the Hamiltonian can be written as
$H(k_{z})|_{k_{x}=k_{y}=0}=d_{0}+d_{1}\sigma_{3}+d_{2}\tau_{3}\sigma_{3}+d_{3}\tau_{3}$
where $d_{0,1,2,3}(k_{z},m)$ are real functions.
Since the simultaneous presence of TRS and IS requires
the double degeneracy of each state, among $d_{1,2,3}$, only
one function can be nonzero.
Also, since the degenerate bands should have the opposite spin directions, $d_{1}=0$.
Hence the Hamiltonian becomes
$H(k_{z})|_{k_{x}=k_{y}=0}=d_{0}+d(k_{z},m)\Gamma$ where $\Gamma$ is either
$\Gamma=\tau_{3}$ or $\Gamma=\sigma_{3}\tau_{3}$.
Then since the energy gap is given by 2$|d(k_{z},m)|$, an ABC
can be achieved if and only if $d(k_{z},m)=0$.
Here the number of variables (two) is larger
than the number of equations (one) to be satisfied for the band crossing,
hence the Dirac SM can always be created via an ABC.

Moreover, due to the TRS and IS,
$d(k_{z})$ has a definite parity under the sign reversal of $k_{z}$ as shown in Methods section.
In fact, the parity of $d(k_{z})$ can be simply determined by the matrix representation $P$ of the IS.
Namely, when $P$ has a diagonal form such as $P=\pm\tau_{0}$ or $\pm\tau_{z}$,
$d(k_{z})$ is even while it is odd if $P$ has an off-diagonal form $P=\pm\tau_{x}$.
At first, when $d(k_{z})$ is even,
$d(k_{z})\approx M+\frac{1}{2}t_{z} k_{z}^{2}$ in the leading order with constants $M$ and $t_{z}$.
In this case, the system is a gapped insulator (a Dirac SM) when $Mt_{z}>0$ ($Mt_{z}<0$).
Namely, by taking $M$ as a tunable parameter and assuming $t_{z}<0$,
the transition from an insulator ($M<0$) to a 3D Dirac SM ($M>0$)
can be achieved across the sign reversal of $M$ (or the band inversion).
In particular, the 3D Dirac SM phase possesses two DPs
which are symmetrically located with respect to the center of the rotation axis
at $k_{z}=\pm\sqrt{2M/|t_{z}|}$.
The Dirac SMs realized in Cd$_{3}$As$_{2}$ and Na$_{3}$Bi belong to this class~\cite{LDA_Wang1,LDA_Wang2}.
On the other hand,
when $d(k_{z})$ is odd, $d(k_{z})\approx \upsilon k_{z}$ in the leading order with a constant $\upsilon$.
In this case, there is a single 3D Dirac point at the center of the $k_{z}$ axis, i.e., at $k_{z}=0$.
Then the system is nothing but a stable 3D Dirac SM with a single DP at the center of the rotation axis.
Considering the full periodic structure of the Brillouin zone (BZ), both $k_{z}=0$ and $k_{z}=\pi$, i.e.,
the TRIMs on the rotation axis,
are the possible locations of the DP.
The candidate Dirac SM systems such as $\beta$-cristobalite BiO$_{2}$~\cite{theory_Young} and distorted spinels~\cite{theory_Young2}
proposed by the recent theoretical studies belong to this class.
In fact, there are multiple
bulk Dirac points in $\beta$-cristobalite BiO$_{2}$ due to the crystalline symmetry.
However, each Dirac point locates at a TRIM on a rotation axis consistent with our theory.
In conclusion, when the material has the TRS, IS, and uniaxial rotation symmetry simultaneously,
there are two different ways to obtain a stable Dirac SM phase.
One is through the transition from an insulator to a Dirac SM via an ABC,
which gives rise to a stable 3D Dirac SM with a pair of 3D bulk DPs on the axis of the rotation.
(See Figure 1c.)
The other case is when the system naturally supports a 3D DP
at a TRIM on the rotation axis due to the symmetry of the system. (See Figure 1b.)
The intrinsic properties of the relevant 3D Dirac SM phases are summarized
in Table 1 and Table 2, respectively.

{\bf Classification table.}
The classification of the 3D Dirac SM phases
can be performed rigorously by imposing the TRS, IS and $C_{n}$
rotation symmetry to the minimal 4$\times$4 Hamiltonian $H(\textbf{k})$.
In the simultaneous presence of the TRS and IS,
the Hamiltonian can always be written as
$H(\textbf{k})=\sum_{i=1}^{5}a_{i}(\textbf{k})\Gamma_{i}$
where $\Gamma_{i}$ indicates 4$\times$4 hermitian matrices
satisfying $\{\Gamma_{i},\Gamma_{j}\}=2\delta_{ij}$,
which guarantees the double degeneracy of eigenstates at each $\textbf{k}$. (See Methods.)
The precise functional form of $a_{i}(\textbf{k})$
can be fixed by imposing the $C_{n}$ rotational symmetry
using the basis under which both $C_{n}$ and $H(\textbf{k})$
are diagonal along the $k_{z}$ direction~\cite{Fang}.
The results of the complete classification are concisely summarized in Table 1 and 2.
Because of the TRS, the
rotation operator becomes $C_{n}=\text{diag}[u_{A,\uparrow},u_{B,\uparrow},u^{*}_{A,\uparrow},u^{*}_{B,\uparrow}]$
where each component $u$ can be written as $u=\exp[i\frac{2\pi}{n}(p+\frac{1}{2})]$
with $p=0,1,...,n-1$ and $A, B$ indicate the orbital degrees of freedom.
Hence only two components of $C_{n}$ are independent.

The properties of the 3D Dirac SM generated via ABCs
are shown in Table 1.
In general, the ABC requires the valence and conduction bands to have different $C_{n}$ eigenvalues.
Namely, $\{u_{A,\uparrow},u_{A,\downarrow}\}$ and
$\{u_{B,\uparrow},u_{B,\downarrow}\}$ should not have common elements
to avoid the interband hybridization.
However, in the case of $C_{2}$ invariant systems, $\{u_{A,\uparrow},u_{A,\downarrow}\}$=
$\{u_{B,\uparrow},u_{B,\downarrow}\}$=$\{i,-i\}$,
hence the ABC is not allowed for the $C_{2}$ invariant systems.
On the other hand, the systems with $C_{3}$, $C_{4}$, and $C_{6}$ symmetries
can support 3D Dirac SM phases.
Although the detailed structure of the Hamiltonian $H(\textbf{k})$ depends
on the specific symmetries of the corresponding system, there are only two different types of low energy effective Hamiltonians
near the bulk gap-closing point.
In the first case, after some suitable unitary transformations,
the effective Hamiltonian near each of the bulk DP can be written as
\begin{displaymath}
H_{\text{Dirac}}(\textbf{q})
\sim
\left( \begin{array}{cc}
q_{x}\tau_{x}+q_{y}\tau_{y}+q_{z}\tau_{z} & 0 \\
0 & -q_{x}\tau_{x}-q_{y}\tau_{y}-q_{z}\tau_{z}
\end{array} \right),
\end{displaymath}
where the momentum $\textbf{q}$ is measured with respect to the DP.
This is the Hamiltonian for the conventional Dirac fermion (the linear Dirac fermion) which is composed of two Weyl fermions
having the Chern number $+1$ or $-1$, respectively.
On the other hand,
in the case of the $C_{6}$ invariant system
with the $\{u_{A,\uparrow},u_{B,\uparrow}\}=\{e^{i\frac{5\pi}{6}},e^{i\frac{\pi}{6}}\}$,
the effective Dirac Hamiltonian is given by
\begin{displaymath}
H_{\text{Dirac}}(\textbf{q})
\sim
\left( \begin{array}{cc}
(q_{x}^{2}-q_{y}^{2})\tau_{x}+2q_{x}q_{y}\tau_{y}+q_{z}\tau_{z} & 0 \\
0 & -(q_{x}^{2}-q_{y}^{2})\tau_{x}-2q_{x}q_{y}\tau_{y}-q_{z}\tau_{z}
\end{array} \right).
\end{displaymath}
Note that this Dirac point is composed of two double Weyl fermions which have
the Chern number $+2$ or $-2$, respectively.~\cite{Fang,doubleweyl} Hence we call this gapless fermions
as the quadratic Dirac fermions.
It is worth to stress that one intriguing common property shared by the 3D Dirac SMs created via ABCs
is that each Dirac SM possesses a quantized 2D topological invariant, which gives rise
to 2D Dirac fermions localized on the surface.
The physical origin of such a nontrivial topological property of the 3D Dirac SM phase
is described in the following section.

The physical properties of the 3D Dirac SM with a Dirac point at the center of the rotation axis
are summarized in Table 2.
Since the IS flips the orbitals in this case ($P=\pm\tau_{x}$),
the doubly degenerate states at each momentum on the $k_{z}$ axis have different orbitals and opposite spin directions.
Therefore the band crossing between degenerate bands requires $\{u_{A,\uparrow},u_{B,\downarrow}\}\cap\{u_{B,\uparrow},u_{A,\downarrow}\}=\emptyset$
contrary to the previous case.
Moreover, due to the additional constraint of $u_{A,\uparrow}=-u_{B,\uparrow}$,
the Dirac SM phase with a single DP cannot exist in the system with $C_{3}$ invariance
while the systems with $C_{2}$, $C_{4}$, $C_{6}$ symmetries
support it. (See Methods.)
Since the DP locates in the $k_{z}=0$ plane, the quantized 2D topological invariant
cannot be defined in the same plane, hence we do not expect any surface state
in this class of the 3D Dirac SMs in general.
One interesting prediction of Table 2 is that when the system has
$C_{6}$ symmetry with $u_{A,\uparrow}=\pm e^{i\frac{3\pi}{6}}$,
the low energy Hamiltonian near the DP can be written as
\begin{displaymath}
H_{\text{Dirac}}(\textbf{q})
\sim
\left( \begin{array}{cc}
(q_{+}^{3}+q_{-}^{3})\tau_{x}+i(q_{+}^{3}-q_{-}^{3})\tau_{y}+q_{z}\tau_{z} & 0 \\
0 & -(q_{+}^{3}+q_{-}^{3})\tau_{x}-i(q_{+}^{3}-q_{-}^{3})\tau_{y}-q_{z}\tau_{z}
\end{array} \right).
\end{displaymath}
Note that this Dirac point is composed of two triple Weyl fermions which have
the Chern number $+3$ or $-3$, respectively.~\cite{Fang} Hence we can call this gapless fermion
as the cubic Dirac fermion.
For the other cases in Table 2, the effective Hamiltonian near the DP is simply described
by the ordinary linear Dirac fermions.

{\bf Topological properties of the 3D Dirac SM.}
One important characteristic of the 3D Dirac SM generated by an ABC
is that it carries a quantized topological invariant although it is a gapless SM.
In fact, the band inversion associated with the ABC is the common origin
of the presence of the 2D topological invariant and the emergence of bulk DPs.
As shown in Table 1, in each case,
a quantized 2D topological invariant can be defined on the $k_{z}=0$ plane
where the ABC occurs.
First of all, since the $k_{z}=0$ plane can be considered as a 2D system
with TRS, a 2D $Z_{2}$ invariant $\nu_{2D}$ is well-defined on it.~\cite{FuKaneMele,MooreBalents} Moreover,
because of the simultaneous presence of the TRS and IS, $\nu_{2D}$ can be determined
by the parities of the occupied bands at the
time reversal invariant momenta.~\cite{Fu_inversion} Therefore when the valence and conduction bands
have the opposite parities ($P=\pm\tau_{z}$),
the band inversion on the $k_{z}=0$ plane, which generates a pair of bulk DPs,
changes $\nu_{2D}$ by 1, i.e., $\Delta\nu_{2D}=1$.
This can be contrasted to the case when two bands have the same parity ($P=\pm\tau_{0}$),
in which $\Delta\nu_{2D}=0$ in spite of the occurrence of the band inversion.

On the other hand, when the system has either $C_{4}$ or $C_{6}$ symmetry,
the $k_{z}=0$ plane carries an integer topological invariant (the mirror Chern number) due to the mirror symmetry of the system.~\cite{TeoFu,Fu_crystallineTI}
Here the mirror symmetry appears due to the simultaneous presence
of the $\pi$ rotation ($R_{\pi}$)
with respect to the $k_{z}$ axis and the IS.
Then the combined operation of the IS and $\pi$ rotation defines the mirror symmetry $M=PR_{\pi}$,
which connects the Hamiltonian $H(k_{x},k_{y},k_{z})$ and $H(k_{x},k_{y},-k_{z})$,
hence the system is invariant under the mirror symmetry ($M$) in the $k_{z}=0$ plane.
Since $M^{2}=-1$, the Hamiltonian can be block-diagonalized
with each block characterized by the mirror eigenvalue $\pm i$, respectively.
Then the Chern number can be defined in each block $(C_{\pm i})$, separately.
Although the total Chern number $C_{+i}+C_{-i}=0$ due to the TRS,
the difference $n_{M}\equiv\frac{1}{2}(C_{i}-C_{-i})$ (the mirror Chern number) can be nonzero.
Note that $R_{\pi}$ exists only in systems with the $C_{4}$ or $C_{6}$ symmetry with
the corresponding $R_{\pi}=C_{4}^{2}$ or $R_{\pi}=C_{6}^{3}$, respectively.
Therefore the $C_{3}$ invariant system is characterized only by the 2D $Z_{2}$ invariant $\nu_{2D}$
while the system with the $C_{4}$ or $C_{6}$ symmetry has both the $Z_{2}$ invariant $\nu_{2D}$
and the mirror Chern number $n_{M}$ where $n_{M}$ and $\nu_{2D}$ are equivalent up to the modulo 2.
In Methods section, we have described how the mirror symmetry manifests in the system
in terms of the effective Hamiltonian $H(\textbf{k})$.

When either $\nu_{2D}$ or $n_{M}$ is nonzero, the 3D Dirac SM supports
2D surface Dirac cones when a surface parallel to the $k_{z}$ axis is introduced.
The number of 2D Dirac cones on one surface is given by $|n_{M}|$ ($|\nu_{2D}|$)
when the system has the $C_{4}$ or $C_{6}$ ($C_{3}$) symmetry.
The typical surface spectrum is composed of two parts.
One is from the 3D bulk Dirac states projected to the surface BZ
and the other is from the 2D surface Dirac cones
resulting from the 2D topological index.
When these two contributions are decoupled on the Fermi level,
the 2D Dirac cone forms an isolated closed loop.
On the other hand, when the bulk and surface states are coupled
on the Fermi level, the 2D surface states form Fermi arcs connecting the bulk states.
The precise shape of the Fermi surface in the surface BZ
depends on the symmetry of the system and detailed material parameters.

It is worth to stress that the physical origin of the surface Fermi arcs
in the 3D Dirac SM is clearly distinct from that of the Weyl SM which has two-fold degeneracy at the gap-closing point.
In the Weyl SM, the Chern number carried by the bulk gapless point (Weyl point)
guarantees the emergence and stability of the Fermi arc states.~\cite{Weyl_iridate, Weyl_Balents,Weyl_William}
Since the Chern number of the Weyl point is purely determined by the energy dispersion around the Weyl point,
the number of Fermi arcs in the system with a fixed number of Weyl points strongly depends
on the energy dispersion near the Weyl point.
In conventional Weyl SMs with the linear dispersion around the Weyl point,
the number of Fermi arcs on one surface of the sample is equal to the number of Weyl point pairs in the first BZ.
On the other hand, in the case of Weyl SMs with double (triple) Weyl fermions
whose dispersion is quadratic (cubic) along the two momentum directions but linear in the third direction,
the number of Fermi arcs is double (triple) of the number of the Weyl point pairs.~\cite{Fang, doubleweyl}
However, in contrast to the case of the Weyl SM, the physical origin of the surface states of the 3D Dirac SM
is independent of the energy dispersion of the 3D bulk Dirac fermions.
Because of the simultaneous presence of the TRS and IS, the Chern number of each 3D Dirac point
is zero, hence the 3D Dirac point is topologically trivial.
Here the number of the Fermi arcs on the surface of the sample is solely determined
by the 2D topological invariant on the $k_{z}=0$ (or $k_{z}=\pi$) plane
irrespective of the energy dispersion around the 3D bulk Dirac points.
Therefore although the low energy Hamiltonian near the 3D bulk Dirac point
is the same, the number of Fermi arcs can be different depending
on the 2D topological invariant of the system.
An example is shown in Figure 3.

{\bf Lattice model and generic phase diagram.}
Let us illustrate the intriguing properties
of the 3D topological Dirac SM phases in Table 1 by studying
lattice Hamiltonians numerically.
For convenience, we choose the $C_{4}$ invariant systems
with $P=\tau_{0}$ (Equation (\ref{eqn:latticeH2}) in Methods.) or $P=\tau_{z}$ (Equation (\ref{eqn:latticeH1}) in Methods.)
corresponding to the 6th or 7th row of Table 1, respectively.
However, the main features of the phase diagram can be applied to all cases in Table 1 because
the overall structure of the phase diagram is solely determined by the single function $a_{5}(\textbf{k})$
whose leading order functional form is the same in all cases.
The detailed information about the lattice Hamiltonian is presented in the Methods section.
Figure 2 summarizes the main properties of the lattice Hamiltonian.
In general, the system supports four different phases as shown in the phase diagram.
The phase transition is always accompanied by an ABC on the $k_{z}=0$ or $k_{z}=\pi$ plane
in which a pair of 3D bulk DPs are either created or annihilated.
Whenever an ABC happens, it changes the 2D topological invariant of the corresponding 2D planes.
When any of these 2D planes has a nonzero topological invariant, the system supports
2D Dirac fermions on the surface which is parallel to the rotation axis.

There are two different types of insulators in the phase diagram.
One is a normal insulator which does not carry a topological number,
and the other is a weak topological insulator in which both
the $k_{z}=0$ and $k_{z}=\pi$ planes have nontrivial 2D topological invariants.
In systems with $P=\pm\tau_{z}$,
the weak topological insulator is equivalent to the conventional weak topological
insulator with the $Z_{2}$ topological index $(\nu_{0};\nu_{1}\nu_{2}\nu_{3})=(0;001)$
because $\nu_{2D}=1$ in both the $k_{z}=0$ and $k_{z}=\pi$ planes.
On the other hand, when $P=\pm\tau_{0}$, the nature of the weak topological insulator in the phase diagram
is unconventional in the sense that
the topological property of the insulator is determined by the nonzero mirror Chern number $n_{M}=2$
on both the $k_{z}=0$ and $k_{z}=\pi$ planes.

The 3D Dirac SM phases can also be distinguished in two different ways.
At first, when the Dirac SM phase
has a nonzero topological invariant in either the $k_{z}=0$ plane or the $k_{z}=\pi$ plane,
we can call it a {\it topological Dirac SM} since the Dirac SM carries stable 2D Dirac cones on the surface.
Similarly, a topologically trivial Dirac SM can be defined
when the system does not have any topological invariant in both planes.
However, in both cases, independent of the presence of the 2D topological invariants,
the 3D Dirac SM phase is stable due to the symmetry of the system
and occupies a finite region in the phase diagram.

Figure 3 shows the evolution of the Fermi surface of the topological Dirac SM system with a slab geometry
whose surface normal is parallel to the [100] direction.
The translational symmetry of the system in the $yz$ plane is maintained.
Here we first pick the states touching the Fermi level ($E_{F}$) and then plot
the wave function amplitudes of the corresponding state localized on the first five layers from the top surface.
The red color indicates the states localized on the [100] surface while
the bright blue color corresponds to the 3D bulk Dirac states.
In the case of the topological Dirac SM with $\nu_{2D}=1$
in the $k_{z}=0$ plane,
a 2D surface Dirac cone appears on the $k_{y}$ axis
centered at the $\Gamma$ point  as shown in Figure 3a.
Here the two bulk Dirac points give rise to the finite intensity on the $k_{z}$ axis located symmetrically
with respect to the $\Gamma$ point.
When the Fermi energy ($E_{F}$) is near the bulk Dirac point ($E_{F}=0$ at the bulk Dirac point),
the bulk and surface states are decoupled, and the surface states
form an isolated closed loop.
As $E_{F}$ increases, the Fermi surface topology evolves continuously, and
when the bulk and surface states start to overlap, the 2D surface state is deformed to
the Fermi arc structure.
Namely, the Fermi arcs of 3D Dirac SM emerge simply because of
the deformation of the 2D Dirac cone which exists due to the fact that $\nu_{2D}=1$ on the $k_{z}=0$ plane.
Since such an evolution of the surface spectrum occurs continuously,
the Fermi arc states can appear even when $E_{F}=0$ if the parameters
of the model Hamiltonian are tuned properly.

The energy spectrum of the topological Dirac SM with $n_{M}=2$ on the $k_{z}=0$ plane
also shows a similar variation as described in Figure 3b.
Since $n_{M}=2$, there are two 2D surface Dirac cones
on the $k_{y}$ axis.
As $E_{F}$ increases, the surface Dirac cones evolve to Fermi arcs
when the surface and bulk states overlap.
Since the number of surface Dirac cones is two, the number of surface Fermi arc
is also doubled as compared to the case shown in Figure 3a.
It is worth to note that in both cases shown in Figure 3a and 3b,
the energy dispersion near the bulk 3D Dirac point is basically the same,
i.e., the bulk state shows the linear dispersion relation in all three momentum directions.
This clearly shows that the number of Fermi arcs of the topological Dirac SM
is irrespective of the dispersion of the bulk states.

\subsection*{Discussion}
Let us illustrate the role of the rotational symmetry by comparing the Figure 1a-c.
When the system has the TRS and IS without the rotational symmetry, corresponding to Figure 1a,
the 3D DP appears at the quantum critical point between the normal insulator
and the strong topological insulator. Since the 3D DP can be observed only at the single point of the phase diagram,
the Dirac SM is not a stable phase in this case.
On the other hand, in the presence of the additional uniaxial rotational symmetry,
two different types of phase diagrams emerge as shown in Figure 1b and 1c.
In both cases, the Dirac SM is a stable phase occupying a finite region of the phase diagram.
At first, due to the rotation symmetry, the Dirac point can persist at a TRIM on the rotation axis independent
of the external control parameter $m$, hence there is only a single phase
in the phase diagram as shown in Figure 1b. The Dirac SMs in Table 2 correspond to this case.
Whereas, when the Dirac SM is generated via an ABC, the phase diagram is
depicted in Figure 1c and the properties of the corresponding Dirac SM are summarized in Table 1.
In fact, the 3D topological Dirac SM phase emerging here
can also be understood as a state intermediating a topological phase transition between two insulators.
Namely, the 3D topological Dirac SM
mediates the transition between a normal insulator and a weak topological insulator.
At the critical points, a pair of 3D DPs are either created ($m=m_{c1}$) or annihilated ($m=m_{c2}$)
at the center or the boundary of the rotation axis, respectively.
Between the two critical points, as $m$ increases, the pair of 3D DPs move along the rotation axis in the opposite directions,
and finally recombine when they hit the BZ boundary.
Since each DP is stable due to the symmetry of the system,
the pair creation or the pair annihilation is the only way to change the number of the DPs
as long as the TRS, IS, and rotation symmetry are present at the same time.

Since the four-fold degeneracy at the DP is protected by symmetries,
it is important to understand the fate of the 3D Dirac SM under symmetry breaking perturbations.
In particular, it is worth to note that the 3D topological Dirac SM itself
can be considered as a parent state of various topological insulators that can be obtained
when the DP acquires a mass gap by breaking the rotation symmetry.
When the topological Dirac SM carries a nontrivial 2D $Z_{2}$ invariant
on either the $k_{z}=0$ or the $k_{z}=\pi$ plane, the Dirac SM
turns into a $Z_{2}$ strong topological insulator as long as
the perturbation does not break the time reversal symmetry.
Also in the case of the topological Dirac SM with a nonzero mirror Chern number,
it becomes a topological crystalline insulator after the gap opening,
as long as the mirror symmetry of the system is not broken
in the presence of the perturbations.

We conclude with a discussion about the stability of the 3D Dirac SM
under the influence of the Coulomb interaction and disorder.
Simple power counting shows that the long range Coulomb
interaction is a marginally irrelevant perturbation
to the 3D Dirac fermions with the linear dispersion.~\cite{Chakravarty,Isobe1,Isobe2}
Hence various physical properties of the 3D Dirac SM can receive logarithmic corrections due to
the long range Coulomb interaction similar to the cases of 3D Weyl SM~\cite{Hosur,Isobe1,Isobe2} and graphene.~\cite{Gonzalez,Kotov_RMP,Son}
On the other hand, since the disorder is irrelevant according to the power counting,
we expect the Dirac SM state can be stable at least against weak disorder effect.
However, since the crystalline symmetry is important for the protection of
the DP, strong disorder can induce nontrivial physical consequences to 3D Dirac SM phase,
especially when the interaction and disorder effect are considered simultaneously.~\cite{Chakravarty}
Moreover,
in the case of the quadratic Dirac SM and the cubic Dirac SM,
the effect of the interaction and disorder can be more significant.
Since the in-plane dispersion becomes either quadratic or cubic in the momentum space,
which strongly enhances
the low energy density of states, it is expected that the interaction and disorder
can even bring about new exotic quantum phases.
For instance, according to a recent theoretical study,
an exotic non-Fermi liquid state can appear in a 3D semimetal having
quadratic energy dispersion in the momentum space.~\cite{Moon}
Since the interplay between the long-range Coulomb interaction and
nontrivial screening due to the enhanced low energy density of states
is the fundamental origin leading to the non-Fermi liquid phase,
the quadratic Dirac SM and the cubic Dirac SM are also promising systems
to observe novel quantum critical states.

Finally, let us note that the quantum critical point in Figure 1c where the pair creation or pair annihilation of
bulk DPs happens is another interesting venue to observe
a new types of quantum critical phenomena. At the quantum critical point, since
the energy dispersion along the rotation axis is always quadratic,
the low energy excitation can show highly anisotropic dispersion relations
in the linear Dirac SM and cubic Dirac SM.
According to the recent theoretical study,~\cite{Yang, Yang_RG} it is shown that
such an anisotropic dispersion can induce a novel screening phenomenon
which can induce anomalous distribution of the screening charge around a charged impurity.
To reveal the fascinating physical properties of the linear Dirac SM and the triple
Dirac SM at the critical point would be another interesting topic for future studies.

\subsection*{Methods}
{\bf The classification procedure.}
The classification of the minimal 4$\times$4 matrix Hamiltonian $H(\textbf{k})$
can be performed as follows.
At first, let us impose the TRS on $H(\textbf{k})$.
The TRS can be represented by the operator $\Theta=i\sigma_{y}K$
where $\sigma_{x,y,z}$ are Pauli matrices for spin degrees of freedom
and $K$ stands for complex conjugation.
The invariance of the Hamiltonian under TRS, i.e.,
$H(-\textbf{k})=\Theta H(\textbf{k}) \Theta^{-1}$ gives
rise to the relations $h_{\uparrow\uparrow}(\textbf{k})=h_{\downarrow\downarrow}^{T}(-\textbf{k})$
and $h_{\uparrow\downarrow}(\textbf{k})=-h_{\uparrow\downarrow}^{T}(-\textbf{k})$
where the superscript $T$ indicates the transposition.
Then the resulting Hamiltonian with the TRS
can be written in the following way.
\begin{displaymath}
H(\textbf{k})=
\left( \begin{array}{cc}
h_{\uparrow\uparrow}(\textbf{k}) & h_{\uparrow\downarrow}(\textbf{k}) \\
-h_{\uparrow\downarrow}^{*}(-\textbf{k}) & h_{\uparrow\uparrow}^{*}(-\textbf{k})
\end{array} \right),
\end{displaymath}
where the superscript $*$ indicates the complex conjugation.
Secondly,
to impose the IS on the Hamiltonian $H(\textbf{k})$,
we have to determine the matrix representation $P$ of the IS.
Since the IS is independent of the spin-rotation, in general $P=p_{0}\tau_{0}+\vec{p}\cdot\vec{\tau}$
where $\vec{\tau}=(\tau_{x},\tau_{y},\tau_{z})$ indicate the Pauli matrices for orbital degrees of freedom
and $p_{0,x,y,z}$ are complex numbers.
Since the operation of $P^{2}$ relates the same electronic states, it should
be equivalent to the identity operator up to a global U(1) phase factor, i.e.,
$P^{2}=p_{0}^{2}+\vec{p}\cdot\vec{p}+2p_{0}\vec{p}\cdot\vec{\tau}=e^{i2\phi}$.
Therefore $P$ should be either $P=\pm e^{i\phi}$ or $P=e^{i\phi}\vec{p'}\cdot\vec{\tau}$
where $\vec{p'}\cdot\vec{p'}=1$.
To determine $\phi$ and $\vec{p'}$, the following three relations can be used.
(i) $[T,P]=0$, (ii) $P^{\dag}P=1$, and (iii) $(TP)^{2}=-1$.
Then the general solution for $P$ is given by
$P=\pm\tau_{0}$ or $P=\cos\theta \tau_{z}-\sin\theta \tau_{x}$ with $\theta\in[0,2\pi]$.

The invariance of the Hamiltonian under $P$, i.e., $H(-\textbf{k})=PH(\textbf{k})P^{-1}$
combined with the TRS constrains the possible form of the Hamiltonian, which can be
summarized in the following way.

(a) When $P=\pm\tau_{0}$.
$H(\textbf{k})=a_{0}(\textbf{k})+\sum_{i=1}^{5}a_{i}(\textbf{k})\Gamma_{i}$
where $\Gamma_{1}=\tau_{x}$, $\Gamma_{2}=\tau_{y}\sigma_{z}$, $\Gamma_{3}=\tau_{y}\sigma_{x}$,
$\Gamma_{4}=\tau_{y}\sigma_{y}$, $\Gamma_{5}=\tau_{z}$.
Hence $h_{\uparrow\uparrow}=a_{0}+a_{1}\tau_{x}+a_{2}\tau_{y}+a_{5}\tau_{z}$ and
$h_{\uparrow\downarrow}=(a_{3}-ia_{4})\tau_{y}$.
Here $a_{0,1,2,3,4,5}(\textbf{k})$ are all real and even under the sign change of $\textbf{k}$.

(b) When $P=\cos\theta \tau_{z}-\sin\theta \tau_{x}$.
$H(\textbf{k})=a_{0}(\textbf{k})+\sum_{i=1}^{5}a_{i}(\textbf{k})\Gamma_{i}$
where $\Gamma_{1}=\mu_{x}\sigma_{z}$, $\Gamma_{2}=\mu_{y}$, $\Gamma_{3}=\mu_{x}\sigma_{x}$,
$\Gamma_{4}=\mu_{x}\sigma_{y}$, $\Gamma_{5}=\mu_{z}$, and all $a_{0,1,2,3,4,5}$ are real functions.
Hence $h_{\uparrow\uparrow}=a_{0}+a_{1}\mu_{x}+a_{2}\mu_{y}+a_{5}\mu_{z}$ and
$h_{\uparrow\downarrow}=(a_{3}-ia_{4})\mu_{x}$.
Here $a_{0,5}(-\textbf{k})=a_{0,5}(\textbf{k})$, $a_{1,2,3,4}(-\textbf{k})=-a_{1,2,3,4}(\textbf{k})$,
and $\mu_{x}=\cos\theta\tau_{x}+\sin\theta\tau_{z}$, $\mu_{y}=\tau_{y}$,
$\mu_{z}=-\sin\theta\tau_{x}+\cos\theta\tau_{z}$.

In both cases,
an ABC is possible only if the five equations $a_{1,2,3,4,5}=0$
are satisfied simultaneously.
Here each function $a_{i}$ has four variables including the three momentum components $k_{x,y,z}$
and one external control parameter $m$, i.e., $a_{i}=a_{i}(k_{x},k_{y},k_{z},m)$.
Since the number of equations to be satisfied is five while the number of variables is four,
the condition for the ABC cannot be satisfied in general at a generic momentum $\textbf{k}$.
However, it is worth to note that the above consideration does not rule out
the ABC at non-generic points in the momentum space
with high symmetry.
For instance, as pointed out by Murakami,~\cite{Murakami0}
at the time reversal invariant momentum $\textbf{k}=\textbf{k}_{\text{TRIM}}$
where $\textbf{k}$ and $-\textbf{k}$ are equivalent,
all odd functions in $H(\textbf{k})$ vanish.
In the case of (b) with $P=\cos\theta \tau_{z}-\sin\theta \tau_{x}$, $a_{1,2,3,4}(\textbf{k}_{\text{TRIM}})=0$.
Therefore an ABC is possible if and only if one condition $a_{5}(\textbf{k}_{\text{TRIM}},m)=0$
is satisfied, which can be achieved by tuning one external control parameter $m$.
This is the reason why the topological phase transition between two insulators
can occur through an ABC at a time reversal invariant momentum.

Now let us show how the condition for ABC is modified by the presence of
the additional rotation symmetry $C_{n}$ with respect to the $z$ axis.
Here $n$ is restricted to be $n=2, 3, 4, 6$ in periodic lattice systems.
Using a basis in which both $H(k_{z})|_{k_{x}=k_{y}=0}$ and $C_{n}$
are diagonal,
$C_{n}$ can be represented by
a diagonal matrix
$C_{n}=\text{diag}[u_{A,\uparrow},u_{B,\uparrow},u_{A,\downarrow},u_{B,\downarrow}]=\text{diag}[\alpha_{p},\alpha_{q},\alpha_{r},\alpha_{s}]$
where $\alpha_{p}=\exp[i\frac{2\pi}{n}(p+\frac{1}{2})]$
with $p=0,1,...,n-1$.~\cite{Fang} For convenience, we express $C_{n}$
in the following way,
\begin{displaymath}
C_{n}=
\left( \begin{array}{cc}
e^{i\pi(\frac{1+p+q}{n}+\frac{p-q}{n}\tau_{z})} & 0 \\
0 & e^{i\pi(\frac{1+r+s}{n}+\frac{r-s}{n}\tau_{z})}
\end{array} \right).
\end{displaymath}
The invariance of
the Hamiltonian under $C_{n}$ leads to
\begin{eqnarray}\label{eqn:Rconstraint}
C_{n}H(k_{+},k_{-},k_{z})C_{n}^{-1}=H(k_{+}e^{i\frac{2\pi}{n}},k_{-}e^{-i\frac{2\pi}{n}},k_{z})
\end{eqnarray}
where $k_{\pm}=k_{x}\pm ik_{y}$.
From Equation (\ref{eqn:Rconstraint}), we can obtain that
\begin{eqnarray}\label{eqn:h_constrained_old}
&e^{i\frac{\pi}{n}(p-q)\tau_{z}}h_{\uparrow\uparrow}(k_{\pm})e^{-i\frac{\pi}{n}(p-q)\tau_{z}}
=h_{\uparrow\uparrow}(k_{\pm}e^{\pm i\frac{2\pi}{n}}),
\nonumber\\
&e^{i\frac{\pi}{n}(p+q-r-s)}e^{i\frac{\pi}{n}(p-q)\tau_{z}}h_{\uparrow\downarrow}(k_{\pm})e^{i\frac{\pi}{n}(s-r)\tau_{z}}
=h_{\uparrow\downarrow}(k_{\pm}e^{\pm i\frac{2\pi}{n}}).
\end{eqnarray}
The equations above can be further simplified by considering
the constraint on $C_{n}$ due to TRS. Namely,
since $[T,C_{n}]=0$, we can show that
$\exp[i\frac{2\pi}{n}(p+r+1)]=1$, $\exp[i\frac{2\pi}{n}(q+s+1)]=1$,
and $C_{n}=\text{diag}[\alpha_{p},\alpha_{q},\alpha_{p}^{*},\alpha_{q}^{*}]$.
Namely, $C_{n}=\text{diag}[u_{A,\uparrow},u_{B,\uparrow},u_{A,\downarrow}=u^{*}_{A,\uparrow},u_{B,\downarrow}=u^{*}_{B,\uparrow}]$.
Then Equation (\ref{eqn:h_constrained_old})
becomes
\begin{eqnarray}\label{eqn:h_constrained}
&e^{i\frac{\pi}{n}(p-q)\tau_{z}}h_{\uparrow\uparrow}(k_{\pm})e^{-i\frac{\pi}{n}(p-q)\tau_{z}}
=h_{\uparrow\uparrow}(k_{\pm}e^{\pm i\frac{2\pi}{n}}),
\nonumber\\
&e^{i\frac{2\pi}{m}(q-r)}e^{i\frac{\pi}{n}(p-q)\tau_{z}}h_{\uparrow\downarrow}(k_{\pm})e^{i\frac{\pi}{n}(p-q)\tau_{z}}
=h_{\uparrow\downarrow}(k_{\pm}e^{\pm i\frac{2\pi}{n}}).
\end{eqnarray}

In general, $h_{\uparrow\uparrow}(\textbf{k})$ and $h_{\uparrow\downarrow}(\textbf{k})$ can be represented by
\begin{eqnarray}
h_{\uparrow\uparrow}(\textbf{k})=f_{0}(\textbf{k})+f_{+}(\textbf{k})\tau_{+}
+f_{+}^{*}(\textbf{k})\tau_{-}+f_{z}(\textbf{k})\tau_{z},
\nonumber\\
h_{\uparrow\downarrow}(\textbf{k})=g_{0}(\textbf{k})+g_{+}(\textbf{k})\tau_{+}
+g_{-}(\textbf{k})\tau_{-}+g_{z}(\textbf{k})\tau_{z},
\end{eqnarray}
where $f_{0,z}$ are real functions while $f_{+}$, $g_{0,\pm,z}$
are complex functions. Also $\tau_{\pm}=\tau_{x}\pm i\tau_{y}$.
Since $f_{0}$ does not affect the gap-closing, we can neglect it in
the forthcoming discussion.
Equation (\ref{eqn:h_constrained}) gives the following relations,
\begin{eqnarray}\label{eqn:constraint_H11}
e^{i\frac{2\pi}{n}(p-q)}f_{+}(k_{\pm},k_{z})&=f_{+}(k_{\pm}e^{\pm i\frac{2\pi}{n}},k_{z}),
\nonumber\\
f_{z}(k_{\pm},k_{z})&=f_{z}(k_{\pm}e^{\pm i\frac{2\pi}{n}},k_{z}),
\end{eqnarray}
and
\begin{eqnarray}\label{eqn:constraint_H12}
e^{i\frac{2\pi}{n}(p-r)}g_{0+z}(k_{\pm},k_{z})&=g_{0+z}(k_{\pm}e^{\pm i\frac{2\pi}{n}},k_{z}),
\nonumber\\
e^{i\frac{2\pi}{n}(q-s)}g_{0-z}(k_{\pm},k_{z})&=g_{0-z}(k_{\pm}e^{\pm i\frac{2\pi}{n}},k_{z}),
\nonumber\\
e^{i\frac{2\pi}{n}(q-r)}g_{\pm}(k_{\pm},k_{z})&=g_{\pm}(k_{\pm}e^{\pm i\frac{2\pi}{n}},k_{z}),
\end{eqnarray}
where $g_{0\pm z}=g_{0}\pm g_{z}$.
The Equations (\ref{eqn:constraint_H11}) and (\ref{eqn:constraint_H12})
are the key results which lead to the full classification of the 3D Dirac SM.

For the classification, we first consider the TRS and IS,
which restricts the possible structure of the Hamiltonian summarized in (a) and (b).
After that the rotational symmetry is imposed to the Hamiltonian by using
the Equations (\ref{eqn:constraint_H11}) and (\ref{eqn:constraint_H12}).
As shown previously, the most general form of $P$
is given by $P=\pm\tau_{0}$ or $P=\cos\theta \tau_{z}-\sin\theta \tau_{x}$ with $\theta\in[0,2\pi]$.
To determine the matrix representation of $P$ and $C_{n}$,
we use the basis in which both the $H(k_{z})|_{k_{x}=k_{y}=0}$ and $C_{n}$
are diagonal.
As noted before, in such a basis, the Hamiltonian should have a diagonal form given by
$H(k_{z})|_{k_{x}=k_{y}=0}=d(k_{z})\Gamma$
with $\Gamma\in\{\tau_{z},\tau_{z}\sigma_{z}\}$.
Since $d(k_{z})$ should have a definite parity under IS as shown in (a) and (b),
the possible form of $P$ is restricted to be $P=\pm\tau_{0}$, $\pm\tau_{x}$, $\pm\tau_{z}$.
In particular, when $P=\pm\tau_{0}$ or $\pm\tau_{z}$,
$d(\textbf{k})$ is even under the sign change of the momentum $\textbf{k}$,
which leads to the 3D topological Dirac SM via an ABC. (Table 1)
On the other hand,  when $P=\pm\tau_{x}$,
$d(\textbf{k})$ is odd under the sign change of $\textbf{k}$,
which gives rise to a 3D Dirac SM with a single bulk Dirac point. (Table 2)

Let us describe the constraints from the rotation symmetry in detail.
When $P=\pm\tau_{0}$,
the Hamiltonian $h_{\uparrow\uparrow}(\textbf{k})$ and $h_{\uparrow\downarrow}(\textbf{k})$
are given by
\begin{eqnarray}
h_{\uparrow\uparrow}(\textbf{k})&=&a_{1}(\textbf{k})\tau_{x}+a_{2}(\textbf{k})\tau_{y}+a_{5}(\textbf{k})\tau_{z},
\nonumber\\
h_{\uparrow\downarrow}(\textbf{k})&=&(a_{3}(\textbf{k})-ia_{4}(\textbf{k}))\tau_{y}.
\end{eqnarray}
On the other hand, when $P=\pm\tau_{z}$,
\begin{eqnarray}
h_{\uparrow\uparrow}(\textbf{k})&=&a_{1}(\textbf{k})\tau_{x}+a_{2}(\textbf{k})\tau_{y}+a_{5}(\textbf{k})\tau_{z},
\nonumber\\
h_{\uparrow\downarrow}(\textbf{k})&=&(a_{3}(\textbf{k})-ia_{4}(\textbf{k}))\tau_{x}.
\end{eqnarray}
In both cases,
from Equations (\ref{eqn:constraint_H11}) and (\ref{eqn:constraint_H12}), we can obtain
\begin{eqnarray}\label{eqn:fullconstraint}
e^{i\frac{2\pi}{n}(p-q)}f_{+}(k_{\pm},k_{z})&=&f_{+}(k_{\pm}e^{\pm i\frac{2\pi}{n}},k_{z}),
\nonumber\\
a_{5}(k_{\pm},k_{z})&=&a_{5}(k_{\pm}e^{\pm i\frac{2\pi}{n}},k_{z}),
\nonumber\\
e^{i\frac{2\pi}{n}(q-r)}g(k_{\pm},k_{z})&=&g(k_{\pm}e^{\pm i\frac{2\pi}{n}},k_{z}),
\end{eqnarray}
where $f_{+}(\textbf{k})=(a_{1}(\textbf{k})-ia_{2}(\textbf{k}))/2$
and $g(\textbf{k})=(a_{3}(\textbf{k})-ia_{4}(\textbf{k}))/2$.
Since $f_{+}$ and $g$ should be zero on the $k_{z}$ axis,
we obtain $e^{i\frac{2\pi}{n}(p-q)}=u_{A,\uparrow}u^{*}_{B,\uparrow}\neq 1$
and $e^{i\frac{2\pi}{n}(q-r)}=u_{B,\uparrow}u^{*}_{A,\downarrow}\neq 1$.
Hence the ABC is possible when $\{u_{A,\uparrow},u_{A,\downarrow}\}\bigcap\{u_{B,\uparrow},u_{B,\downarrow}\}=\emptyset$,
i.e., when the valence and conduction bands have no rotation eigenvalue in common.

Finally, when $P=\pm\tau_{x}$,
the Hamiltonian $h_{\uparrow\uparrow}(\textbf{k})$ and $h_{\uparrow\downarrow}(\textbf{k})$
are given by
\begin{eqnarray}
h_{\uparrow\uparrow}(\textbf{k})&=&a_{5}(\textbf{k})\tau_{x}+a_{2}(\textbf{k})\tau_{y}+a_{1}(\textbf{k})\tau_{z},
\nonumber\\
h_{\uparrow\downarrow}(\textbf{k})&=&(a_{3}(\textbf{k})-ia_{4}(\textbf{k}))\tau_{z}.
\end{eqnarray}
From the Equations (\ref{eqn:constraint_H11}) and (\ref{eqn:constraint_H12}), we can obtain
\begin{eqnarray}\label{eqn:fullconstraint_taux}
e^{i\frac{2\pi}{n}(p-q)}f_{+}(q_{\pm},k_{z})&=&f_{+}(q_{\pm}e^{\pm i\frac{2\pi}{n}},k_{z}),
\nonumber\\
a_{1}(q_{\pm},k_{z})&=&a_{1}(q_{\pm}e^{\pm i\frac{2\pi}{n}},k_{z}),
\nonumber\\
e^{i\frac{2\pi}{n}(p-r)}g_{z}(q_{\pm},k_{z})&=&e^{i\frac{2\pi}{n}(q-s)}g_{z}(q_{\pm},k_{z})
=g_{z}(q_{\pm}e^{\pm i\frac{2\pi}{n}},k_{z}),
\end{eqnarray}
where $f_{+}(\textbf{k})=(a_{5}(\textbf{k})-ia_{2}(\textbf{k}))/2$
and $g_{z}(\textbf{k})=(a_{3}(\textbf{k})-ia_{4}(\textbf{k}))$.
From the condition that $f_{+}=g_{z}=0$ on the $k_{z}$ axis,
we obtain $e^{i\frac{2\pi}{n}(p-q)}=u_{A,\uparrow}u^{*}_{B,\uparrow}\neq 1$,
$e^{i\frac{2\pi}{n}(p-r)}=u_{A,\uparrow}u^{*}_{A,\downarrow}\neq 1$,
$e^{i\frac{2\pi}{n}(q-s)}=u_{B,\uparrow}u^{*}_{B,\downarrow}\neq 1$,
and $u^{2}_{A,\uparrow}=u^{2}_{B,\uparrow}$.
Namely, (i) $\{u_{A,\uparrow},u_{B,\downarrow}\}\bigcap\{u_{B,\uparrow},u_{A,\downarrow}\}=\emptyset$
and (ii) $u_{A,\uparrow}=-u_{B,\uparrow}$ are the two conditions to be satisfied.
The physical meaning of these two conditions is as follows.
First of all, since the IS flips the orbitals in this case ($P=\pm\tau_{x}$), $E_{A,\sigma}(k_{z})=E_{B,\sigma}(-k_{z})$.
Then the combined operation
of the TRS and IS ensures $E_{A,\uparrow}(k_{z})=E_{B,\downarrow}(k_{z})$
and $E_{A,\downarrow}(k_{z})=E_{B,\uparrow}(k_{z})$ on the $k_{z}$ axis.
Note that the orbital index is physically meaningful in this case,
because the angular momentum is a good quantum number on the $k_{z}$ axis.
Then the crossing between two degenerate bands requires $\{u_{A,\uparrow},u_{B,\downarrow}\}\cap\{u_{B,\uparrow},u_{A,\downarrow}\}=\emptyset$.
Moreover, since $(C_{n}P)^{2}=(PC_{n})^{2}$ on the $k_{z}$ axis,
$C_{n}P=\pm PC_{n}$. However, since the condition (i) is violated if $[C_{n},P]=0$ is fulfilled,
we obtain $\{C_{n},P\}=0$, which immediately leads to the condition (ii).
In fact, the condition (ii) $u_{A,\uparrow}=-u_{B,\uparrow}$ means
$\exp[i\frac{2\pi}{n}(p+\frac{1}{2})]=-\exp[i\frac{2\pi}{n}(q+\frac{1}{2})]$
with integers $p$ and $q$. But this relation cannot be satisfied if $n=3$.
Therefore the 3D Dirac SM with a single DP cannot exist in systems with $C_{3}$
invariance.

{\bf The emergence of the mirror symmetry in the Hamiltonian.}
Let us briefly describe how the mirror symmetry manifests in the effective Hamiltonian
$H(\textbf{k})=\sum_{i=1}^{5}a_{i}(\textbf{k})\Gamma_{i}$.
As shown in Table 1, in systems with the $C_{4}$ or $C_{6}$ symmetry,
either $f(\textbf{k})$ or $g(\textbf{k})$ becomes zero in the $k_{z}=0$ plane.
Although only the lowest order terms are shown in Table 1, we can show
that the same result holds in all orders. This means that
the effective Hamiltonian can be written as $H(\textbf{k})=\sum_{i=1,2,3}a'_{i}(\textbf{k})\Gamma'_{i}$
in the $k_{z}=0$ plane. Here $\Gamma'_{1,2,3}$ are three mutually anti-commuting Gamma matrices.
Since only three Gamma matrices appear in the Hamiltonian, we can define
a conserved quantity $\Gamma'_{4}\Gamma'_{5}$ satisfying $[H(\textbf{k}),\Gamma'_{4}\Gamma'_{5}]=0$.
It is straightforward to show that $\Gamma'_{4}\Gamma'_{5}$ is equivalent to
the mirror operator $M$ in all cases,
hence the system has the mirror symmetry in the $k_{z}=0$ plane.

{\bf The lattice Hamiltonians.}
We can construct the lattice Hamiltonians straightforwardly
by using the information in Table 1 and 2.
For instance, for the $C_{4}$ invariant system with $P=\pm\tau_{z}$, i.e., 6-th row of Table 1,
we can use
\begin{eqnarray}
f&=&\eta(\sin k_{x} + i\sin k_{y}),
\nonumber\\
g&=&\sin k_{z}[(\beta+\gamma)(\cos k_{y}-\cos k_{x})+i(\beta-\gamma)\sin k_{x} \sin k_{y}],
\nonumber\\
a_{5}&=& M-t_{xy}(\cos k_{x}+\cos k_{y})-t_{z}\cos k_{z},
\end{eqnarray}
where $\eta$, $\beta$, $\gamma$, $M$, $t_{xy}$, $t_{z}$ are real constants.
More explicitly,
\begin{eqnarray}
H&=&\sum_{\textbf{k}}\eta[\sin k_{x}c^{\dag}(\textbf{k})\tau_{x}\sigma_{z}c(\textbf{k}) -\sin k_{y}c^{\dag}(\textbf{k})\tau_{y}c(\textbf{k})]
\nonumber\\
&+&\sum_{\textbf{k}}(\beta+\gamma)\sin k_{z}(\cos k_{y}-\cos k_{x})[c^{\dag}(\textbf{k})\tau_{x}\sigma_{x}c(\textbf{k})]
\nonumber\\
&+&\sum_{\textbf{k}}(-1)(\beta-\gamma)\sin k_{z}\sin k_{x} \sin k_{y} [c^{\dag}(\textbf{k})\tau_{x}\sigma_{y}c(\textbf{k})]
\nonumber\\
&+&\sum_{\textbf{k}}[M-t_{xy}(\cos k_{x}+\cos k_{y})-t_{z}\cos k_{z}][c^{\dag}(\textbf{k})\tau_{z}c(\textbf{k})],
\end{eqnarray}
where $c^{\dag}=[c^{\dag}_{A,\uparrow},c^{\dag}_{B,\uparrow},c^{\dag}_{A,\downarrow},c^{\dag}_{B,\downarrow}]$.
In real space, the Hamiltonian becomes
\begin{eqnarray}\label{eqn:latticeH1}
H&=&\frac{\eta}{2}\sum_{n}[-ic^{\dag}_{n}\tau_{x}\sigma_{z}c_{n+\hat{x}}+ic^{\dag}_{n}\tau_{y}c_{n+\hat{y}}+h.c.]
\nonumber\\
&+&\frac{(\beta+\gamma)}{4}\sum_{n}[-ic^{\dag}_{n}\tau_{x}\sigma_{x}c_{n+\hat{y}+\hat{z}}-ic^{\dag}_{n}\tau_{x}\sigma_{x}c_{n-\hat{y}+\hat{z}}
+ic^{\dag}_{n}\tau_{x}\sigma_{x}c_{n+\hat{x}+\hat{z}}+ic^{\dag}_{n}\tau_{x}\sigma_{x}c_{n-\hat{x}+\hat{z}}
+h.c.]
\nonumber\\
&+&\frac{(\beta-\gamma)}{8}\sum_{n}[-ic^{\dag}_{n}\tau_{x}\sigma_{y}c_{n+\hat{x}+\hat{y}+\hat{z}}-ic^{\dag}_{n}\tau_{x}\sigma_{y}c_{n-\hat{x}-\hat{y}+\hat{z}}
-ic^{\dag}_{n}\tau_{x}\sigma_{y}c_{n+\hat{x}-\hat{y}-\hat{z}}-ic^{\dag}_{n}\tau_{x}\sigma_{y}c_{n-\hat{x}+\hat{y}-\hat{z}}+h.c.]
\nonumber\\
&+&M\sum_{n}c^{\dag}_{n}\tau_{z}c_{n}
-\frac{t_{xy}}{2}\sum_{n}[c^{\dag}_{n}\tau_{z}c_{n+\hat{x}}+c^{\dag}_{n}\tau_{z}c_{n+\hat{y}}+h.c.]
-\frac{t_{z}}{2}\sum_{n}[c^{\dag}_{n}\tau_{z}c_{n+\hat{z}}+h.c.],
\end{eqnarray}
where $n$ indicates the lattice sites and $\hat{x}, \hat{y}, \hat{z}$ are
the unit lattice vectors along $x, y, z$ directions.
$\eta$ indicates the nearest neighbor hopping amplitudes in the $xy$ plane,
$(\beta+\gamma)$ denotes the next nearest neighbor hopping amplitudes in the $yz$ and $zx$ planes,
and $(\beta-\gamma)$ indicates the hopping process along the body-diagonal direction of the cubic lattice.
$M$ indicates the on-site potential difference between the $A$ and $B$ orbitals,
and $t_{xy}$ ($t_{z}$) describes the hopping amplitude difference in different orbitals along the $x$, $y$
directions (in the $z$ direction).

Similarly, for the $C_{4}$ invariant system with $P=\pm\tau_{0}$, i.e., 7-th row of Table 1,
we use
\begin{eqnarray}
f&=&\eta\sin k_{z}(\sin k_{x} + i\sin k_{y}),
\nonumber\\
g&=&[(\beta+\gamma)(\cos k_{y}-\cos k_{x})+i(\beta-\gamma)\sin k_{x} \sin k_{y}],
\nonumber\\
a_{5}&=& M-t_{xy}(\cos k_{x}+\cos k_{y})-t_{z}\cos k_{z}.
\end{eqnarray}
More explicitly,
\begin{eqnarray}
H&=&\sum_{\textbf{k}}\eta[\sin k_{x}\sin k_{z}c^{\dag}(\textbf{k})\tau_{x}c(\textbf{k}) -\sin k_{y}\sin k_{z}c^{\dag}(\textbf{k})\tau_{y}\sigma_{z}c(\textbf{k})]
\nonumber\\
&+&\sum_{\textbf{k}}(\beta+\gamma)(\cos k_{y}-\cos k_{x})[c^{\dag}(\textbf{k})\tau_{y}\sigma_{x}c(\textbf{k})]
\nonumber\\
&+&\sum_{\textbf{k}}(-1)(\beta-\gamma)\sin k_{x} \sin k_{y} [c^{\dag}(\textbf{k})\tau_{y}\sigma_{y}c(\textbf{k})]
\nonumber\\
&+&\sum_{\textbf{k}}[M-t_{xy}(\cos k_{x}+\cos k_{y})-t_{z}\cos k_{z}][c^{\dag}(\textbf{k})\tau_{z}c(\textbf{k})],
\end{eqnarray}
In real space, the Hamiltonian becomes
\begin{eqnarray}\label{eqn:latticeH2}
H&=&-\frac{\eta}{4}\sum_{n}[c^{\dag}_{n}\tau_{x}c_{n+\hat{x}+\hat{z}}-c^{\dag}_{n}\tau_{x}c_{n-\hat{x}+\hat{z}}
-c^{\dag}_{n}\tau_{y}\sigma_{z}c_{n+\hat{y}+\hat{z}}+c^{\dag}_{n}\tau_{y}\sigma_{z}c_{n-\hat{y}+\hat{z}}+h.c.]
\nonumber\\
&-&\frac{(\beta+\gamma)}{2}\sum_{n}[c^{\dag}_{n}\tau_{y}\sigma_{x}c_{n+\hat{x}}-c^{\dag}_{n}\tau_{y}\sigma_{x}c_{n+\hat{y}}+h.c.]
\nonumber\\
&+&\frac{(\beta-\gamma)}{4}\sum_{n}[c^{\dag}_{n}\tau_{y}\sigma_{y}c_{n+\hat{x}+\hat{y}}-c^{\dag}_{n}\tau_{y}\sigma_{y}c_{n+\hat{x}-\hat{y}}+h.c.]
\nonumber\\
&+&M\sum_{n}c^{\dag}_{n}\tau_{z}c_{n}
-\frac{t_{xy}}{2}\sum_{n}[c^{\dag}_{n}\tau_{z}c_{n+\hat{x}}+c^{\dag}_{n}\tau_{z}c_{n+\hat{y}}+h.c.]
-\frac{t_{z}}{2}\sum_{n}[c^{\dag}_{n}\tau_{z}c_{n+\hat{z}}+h.c.],
\end{eqnarray}
where $(\beta+\gamma)$ indicates the nearest neighbor hopping amplitudes in the $xy$ plane,
$(\beta-\gamma)$ $(\eta)$ denotes the next nearest neighbor hopping amplitudes in the $xy$ ($yz$ and $zx$) planes.
$M$, $t_{xy}$, $t_{z}$ have the same meaning as above.
In both cases, we have chosen $\eta=1$, $\beta=2$, $\gamma=1$, $t_{z}=1$
while varying $M$ and $t_{xy}$ for the numerical computation.

\bibliographystyle{naturemag}

{\small \subsection*{Acknowledgements}
We greatly appreciate the stimulating discussion with Xi Dai.
We are grateful for support from the Japan Society for the Promotion of Science (JSPS)
through the `Funding Program for World-Leading Innovative R\&D on Science and Technology (FIRST Program)}

\newpage

\begin{table*}[h]
\begin{tabular}{c c c c c c c c c c c c c c c}
\hline
\hline
$C_{n}$ & & $|P|$& &$(u_{A,\uparrow},u_{B,\uparrow})$ & & f($k_{\pm}$, $k_{z}$) & & g($k_{\pm}$, $k_{z}$)  & & 2D topological invariant & & $H_{\text{Dirac}}(\textbf{q})$ && Materials\\
\hline
\hline
$C_{2}$ & & $\tau_{z}$ & & $-$ & & $-$ & & $-$ & & $-$  & & Not allowed & &\\
$C_{2}$ & & $\tau_{0}$ & & $-$ & & $-$ & & $-$ & & $-$  & & Not allowed & &\\
\hline
\hline
$C_{3}$ & & $\tau_{z}$ & &$(e^{i\pi},e^{i\frac{\pi}{3}})$ & & $\beta k_{+}$ & & $\gamma k_{-}$  & & $\nu_{2D}=1$ & & Linear Dirac & & Na$_{3}$Bi~\cite{LDA_Wang1}\\
$C_{3}$ & & $\tau_{0}$ & &$(e^{i\pi},e^{i\frac{\pi}{3}})$ & & $\beta k_{z}k_{+}+\gamma k_{-}^{2}$ & & $\eta k_{z}k_{-}+\xi k_{+}^{2}$  & &$\nu_{2D}=0$ & & Linear Dirac& &\\
\hline
\hline
$C_{4}$ & & $\tau_{z}$ & &$(e^{i\frac{3\pi}{4}},e^{i\frac{\pi}{4}})$ & & $\eta k_{+}$ & & $\beta k_{z}k_{+}^{2}+\gamma k_{z}k_{-}^{2}$  & & $n_{M}=\pm1$ & & Linear Dirac
& & Cd$_{3}$As$_{2}$~\cite{LDA_Wang2}\\
$C_{4}$ & & $\tau_{0}$ & &$(e^{i\frac{3\pi}{4}},e^{i\frac{\pi}{4}})$ & & $\eta k_{z}k_{+}$ & & $\beta k_{+}^{2}+\gamma k_{-}^{2}$  & &
$n_{M}=2\text{sgn}(|\beta|-|\gamma|)$ & & Linear Dirac & &\\
\hline
\hline
$C_{6}$ & & $\tau_{z}$ & &$(e^{i\frac{\pi}{2}},e^{i\frac{\pi}{6}})$ & & $\beta k_{+}$ & & $\gamma k_{z}k_{+}^{2}$ & & $n_{M}=\pm1$ & & Linear Dirac & &\\
$C_{6}$ & & $\tau_{0}$ & &$(e^{i\frac{\pi}{2}},e^{i\frac{\pi}{6}})$ & & $\beta k_{z}k_{+}$ & & $\gamma k_{+}^{2}$ & & $n_{M}=\pm2$ & & Linear Dirac & &\\
\hline
$C_{6}$ & & $\tau_{z}$ & &$(e^{i\frac{5\pi}{6}},e^{i\frac{\pi}{2}})$ & & $\beta k_{+}$ & & $\gamma k_{z}k_{-}^{2}$ & & $n_{M}=\pm1$ & & Linear Dirac & &\\
$C_{6}$ & & $\tau_{0}$ & &$(e^{i\frac{5\pi}{6}},e^{i\frac{\pi}{2}})$ & & $\beta k_{z}k_{+}$ & & $\gamma k_{-}^{2}$ & & $n_{M}=\pm2$ & & Linear Dirac & &\\
\hline
$C_{6}$ & & $\tau_{z}$ & &$(e^{i\frac{5\pi}{6}},e^{i\frac{\pi}{6}})$ & & $\eta k_{z}k_{+}^{2}$ & &$\beta k_{+}^{3}+\gamma k_{-}^{3}$ & &
$n_{M}=3\text{sgn}(|\beta|-|\gamma|$) & & Quadratic Dirac & &\\
$C_{6}$ & & $\tau_{0}$ & &$(e^{i\frac{5\pi}{6}},e^{i\frac{\pi}{6}})$ & & $\eta k_{+}^{2}$ & &$\beta k_{z}k_{+}^{3}+\gamma k_{z}k_{-}^{3}$ & & $n_{M}=\pm2$
& & Quadratic Dirac & &\\
\hline \hline
\end{tabular}
\end{table*}
{\noindent {\bf Classification table for 3D topological Dirac semimetals.}
{
Classification table for 3D topological Dirac semimetals
obtained by an accidental band crossing in systems having $C_{n}$ rotational
symmetry with respect to the $z$ axis. Here
$C_{n}=\text{diag}[u_{A,\uparrow},u_{B,\uparrow},u_{A,\downarrow}=u^{*}_{A,\uparrow},u_{B,\downarrow}=u^{*}_{B,\uparrow}]$
and $\beta$, $\gamma$, $\eta$, $\xi$ are complex numbers.
For compact presentation, $u_{A,\uparrow}$ and $u_{B,\uparrow}$
are arranged in a way that $0<\text{arg}(u_{B,\uparrow})<\text{arg}(u_{A,\uparrow})\leq\pi$.
$\nu_{2D}$ ($n_{M}$) indicates the 2D $Z_{2}$ invariant (mirror Chern number) defined on
the $k_{z}=0$ plane. ($n_{M}=\nu_{2D}$ mod 2.)
The $2\times 2$ Hamiltonian
$h_{\uparrow\uparrow}(\textbf{k})=f(\textbf{k})\tau_{+}+f^{*}(\textbf{k})\tau_{-}+a_{5}(\textbf{k})\tau_{z}$.
In the case of $h_{\uparrow\downarrow}(\textbf{k})$, $h_{\uparrow\downarrow}(\textbf{k})=g(\textbf{k})\tau_{x}$ when $P=\pm\tau_{z}$
while $h_{\uparrow\downarrow}(\textbf{k})=g(\textbf{k})\tau_{y}$ when $P=\pm\tau_{0}$.
The leading order terms of $f(\textbf{k})$ and $g(\textbf{k})$ are shown in the table.
$H_{\text{Dirac}}(\textbf{q})$ describes the effective Hamiltonian near the bulk Dirac point,
which is either $H_{\text{Dirac}}(\textbf{q})=\upsilon_{x}q_{x}\Gamma_{1}+\upsilon_{y}q_{y}\Gamma_{2}+\upsilon_{z}q_{z}\Gamma_{3}$ (Linear Dirac)
or $H_{\text{Dirac}}(\textbf{q})=\upsilon_{x}(q^{2}_{x}-q^{2}_{y})\Gamma_{1}+2\upsilon_{y}q_{x}q_{y}\Gamma_{2}+\upsilon_{z}q_{z}\Gamma_{3}$ (Quadratic Dirac)
where $\Gamma_{1,2,3}$ are mutually anticommuting $4\times 4$
Gamma matrices and $\upsilon_{x,y,z}$ are real constants.
Here the momentum $\textbf{q}$ is measured with respect to the bulk Dirac point.
}
}

\newpage

\begin{table*}[h]
\begin{tabular}{c c c c c c c c c c c c c}
\hline
\hline
$C_{n}$ & & $|P|$& &$u_{A,\uparrow}$ & & f($k_{\pm}$, $k_{z}$) & & $g_{z}$($k_{\pm}$, $k_{z}$)   & & $H_{\text{Dirac}}(\textbf{q})$ && Material\\
\hline
\hline
$C_{2}$ & & $\tau_{x}$ & & $e^{i\frac{\pi}{2}}$ & & $k_{z}F_{1}^{(1)}(k_{x,y})-iF_{2}^{(1)}(k_{x,y})$
& & $\alpha k_{x}+\beta k_{y}$   & & Linear Dirac && Distorted Spinels~\cite{theory_Young2}\\
\hline
\hline
$C_{3}$ & & $\tau_{x}$ & &$-$ & & $-$ & & $-$   & & Not allowed &&\\
\hline
\hline
$C_{4}$ & & $\tau_{x}$ & &$e^{\pm i\frac{\pi}{4}}$ & & $F_{1}^{(2)}(k_{x,y})-ik_{z}F_{2}^{(2)}(k_{x,y})$
& & $\alpha k_{\pm}$   & & Linear Dirac&& BiO$_{2}$~\cite{theory_Young}\\
\hline
\hline
$C_{6}$ & & $\tau_{x}$ & &$e^{\pm i\frac{\pi}{6}}$ & & $k_{z}F_{1}^{(3)}(k_{x,y})
+iF_{2}^{(3)}(k_{x,y})$ & & $\alpha k_{\pm}$  & & Linear Dirac &&\\
\hline
$C_{6}$ & & $\tau_{x}$ & &$e^{i\frac{3\pi}{6}}$ & & $k_{z}F_{1}^{(3)}(k_{x,y})
+iF_{2}^{(3)}(k_{x,y})$ & &$F_{3}^{(3)}(k_{x,y})
+iF_{4}^{(3)}(k_{x,y})$
& & Cubic Dirac &&\\
\hline \hline
\end{tabular}
\end{table*}
{\noindent {\bf Classification table for 3D Dirac SMs with a Dirac point at a TRIM on the rotation axis.}
{
Classification table for 3D topological Dirac semimetals in systems having $C_{n}$ rotational
symmetry with respect to the $z$ axis when $P=\pm\tau_{x}$.
In this Dirac SM phase, the location of the 3D Dirac point is fixed either at the center or the edge of the rotation axis,
i.e., at a TRIM on the rotation axe.
Here $C_{n}=\text{diag}[u_{A,\uparrow},u_{B,\uparrow},u_{A,\downarrow},u_{B,\downarrow}]$
$=\text{diag}[u_{A,\uparrow},-u_{A,\uparrow},u^{*}_{A,\uparrow},-u^{*}_{A,\uparrow}]$
and $\alpha$, $\beta$ are complex numbers.
For compact presentation, $\text{arg}(u_{A,\uparrow})$ is fixed to be
$-\frac{\pi}{2}\leq\text{arg}(u_{A,\uparrow})\leq\frac{\pi}{2}$.
But the same result holds even if $\text{arg}(u_{A,\uparrow})$ is shifted by $\pi$.
The real functions $F^{(1,2,3)}$ are given by
$F_{i=1,2}^{(1)}=c^{(1)}_{i}k_{x}+d^{(1)}_{i}k_{y}$,
$F_{i=1,2}^{(2)}=c^{(2)}_{i}(k_{x}^{2}+k_{y}^{2})+d^{(2)}_{i}k_{x}k_{y}$,
$F_{i=1,2,3,4}^{(3)}=c^{(3)}_{i}(k_{+}^{3}+k_{-}^{3})+id^{(3)}_{i}(k_{+}^{3}-k_{-}^{3})$
where $c^{(1,2,3)}_{i}$ and $d^{(1,2,3)}_{i}$ are real constants.
The $2\times 2$ Hamiltonian
$h_{\uparrow\uparrow}(\textbf{k})=f(\textbf{k})\tau_{+}+f^{*}(\textbf{k})\tau_{-}+a_{1}(\textbf{k})\tau_{z}$
where $a_{1}(\textbf{k})=\upsilon k_{z}$ with a real constant $\upsilon$,
and $h_{\uparrow\downarrow}(\textbf{k})=g_{z}(\textbf{k})\tau_{z}$.
The leading order terms of $f(\textbf{k})$ and $g_{z}(\textbf{k})$ are shown in the table.
$H_{\text{Dirac}}(\textbf{q})$ describes the effective Hamiltonian near the bulk Dirac point,
which is either $H_{\text{Dirac}}(\textbf{q})=\upsilon_{x}q_{x}\Gamma_{1}+\upsilon_{y}q_{y}\Gamma_{2}+\upsilon_{z}q_{z}\Gamma_{3}$ (Linear Dirac)
or $H_{\text{Dirac}}(\textbf{q})=\upsilon_{x}(q^{3}_{+}+q^{3}_{-})\Gamma_{1}+i\upsilon_{y}(q_{+}^{3}-q_{-}^{3})\Gamma_{2}+\upsilon_{z}q_{z}\Gamma_{3}$ (Cubic Dirac)
where the momentum $\textbf{q}$ is measured with respect to the bulk Dirac point with $q_{\pm}=q_{x}\pm iq_{y}$.
Here $\Gamma_{1,2,3}$ are mutually anticommuting $4\times 4$
Gamma matrices and $\upsilon_{x,y,z}$ are real constants.
}
}

\begin{figure*}[t]
\centering
\includegraphics[width=16 cm]{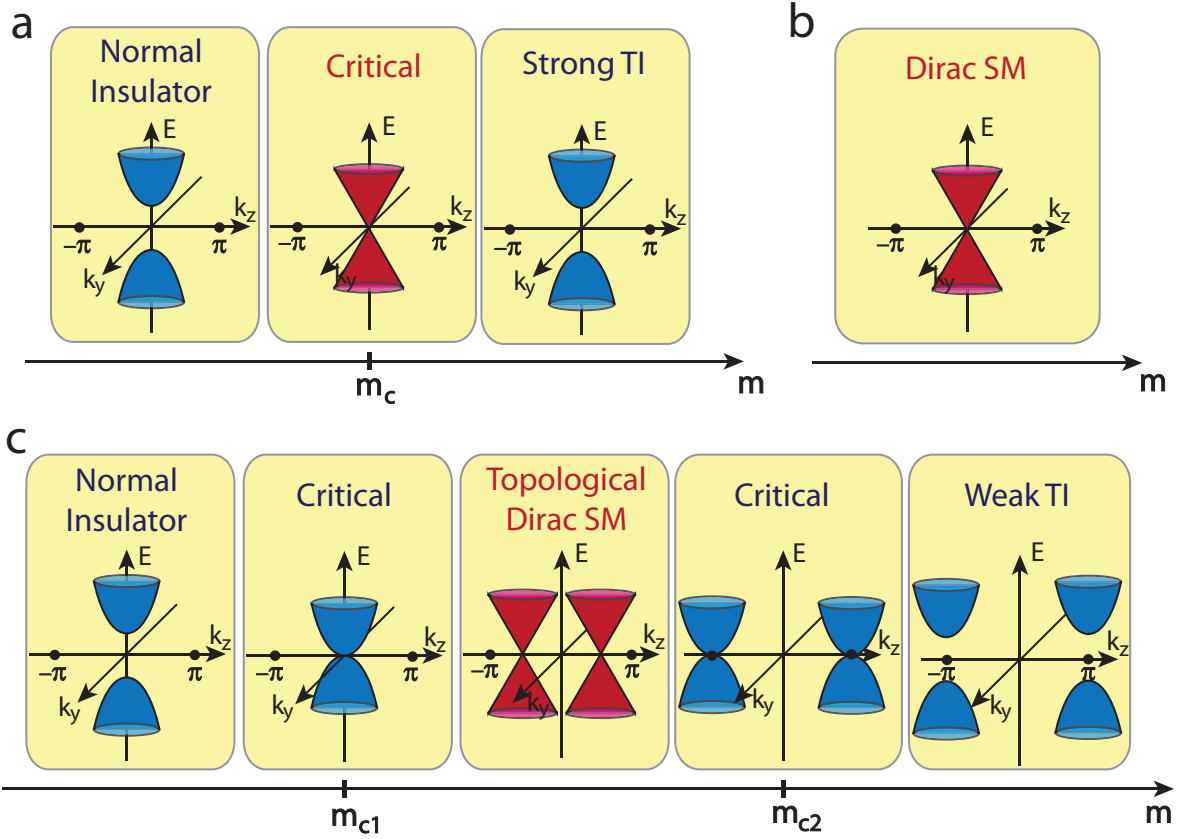}
\caption{
{\bf Creation of a topological Dirac semimetal (SM) via accidental gap closings.}
({\bf a}) The phase transition induced by an ABC when a single control parameter $m$ is varied
in systems with the TRS and IS but lacking
the rotation symmetry. The 3D Dirac fermion appears only at the critical point $m=m_{c}$
and the ABC mediates the transition between a normal insulator
and a strong topological insulator (TI).
When an additional uniaxial rotational symmetry is included,
two different phase diagrams can be obtained as shown in ({\bf b}) and ({\bf c}).
Here we choose the $k_{z}$ axis as the axis for the $n$-fold rotation.
({\bf b}) The Dirac SM persists irrespective of $m$.
The Dirac point locates at a TRIM on the rotation axis.
The Dirac SM in Table 2 corresponds to this case.
({\bf c})
A stable topological Dirac SM phase appears when $m_{c1}<m<m_{c2}$,
which mediates the transition between a normal insulator and a weak TI.
Here a pair of bulk Dirac points, each of which has four-fold degeneracy at the gapless point,
exist along the rotation axis and approach the Brillouin zone (BZ)
boundary as $m$ increases.
The Dirac SM in Table 1 corresponds to this case.
At the quantum critical points ($m=m_{c1}$ or $m=m_{c2}$),
the energy dispersion along the $k_{z}$ direction is quadratic while the dispersion
along the $k_{x}$ and $k_{y}$ directions is linear (linear Dirac SM) or quadratic (quadratic Dirac SM)
or cubic (cubic Dirac SM).
} \label{fig:bandcrossing}
\end{figure*}

\begin{figure*}[t]
\centering
\includegraphics[width=16 cm]{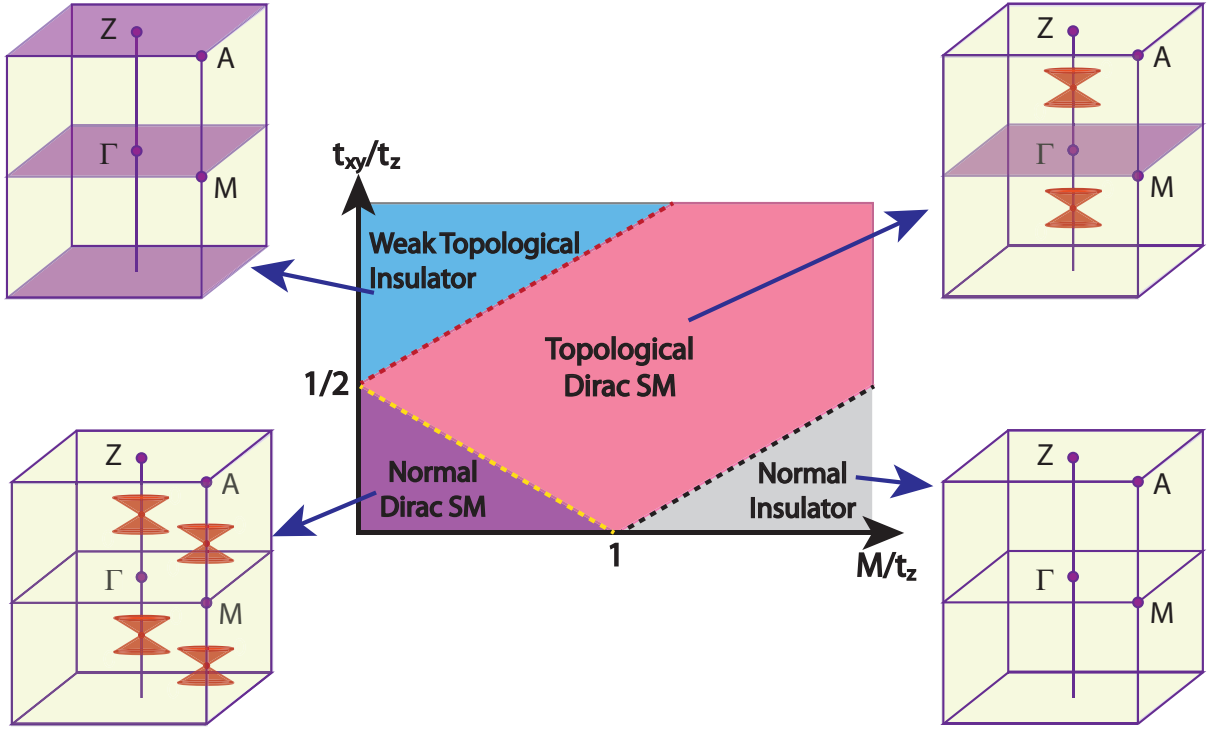}
\caption{
{\bf Generic phase diagram of the system with the time-reversal, inversion, and rotation symmetries.}
The phase diagram is obtained by numerically solving
the lattice Hamiltonians in Equations (\ref{eqn:latticeH1}) and (\ref{eqn:latticeH2}),
both of which lead to the same phase diagram.
Here $M$ indicates the on-site energy difference between two orbitals and
$t_{xy}$ ($t_{z}$) describes the hopping amplitude along the direction perpendicular (parallel)
to the rotation axis.
For each phase, the location of the 3D bulk Dirac point in the first BZ, if it is present, is indicated
by the red symbol having the shape of a Dirac cone.
When the two dimensional (2D) plane with $k_{z}=0$ or $k_{z}=\pi$
possesses a nontrivial 2D topological invariant, such as $Z_{2}$ invariant ($\nu_{2D}$) or the mirror Chern number ($n_{M}$),
the corresponding plane is colored in purple.
If any of these 2D planes carries a nonzero 2D topological invariant,
the surface of the material, which is parallel to the axis of the rotation, supports
2D surface Dirac cones.
A gap-closing happens at the $\Gamma$ ($Z$) point on the black (red) dotted line
while a gap-closing occurs at the $M$ point on the yellow dotted line.
} \label{fig:latticemodel}
\end{figure*}

\begin{figure*}[t]
\centering
\includegraphics[width=16 cm]{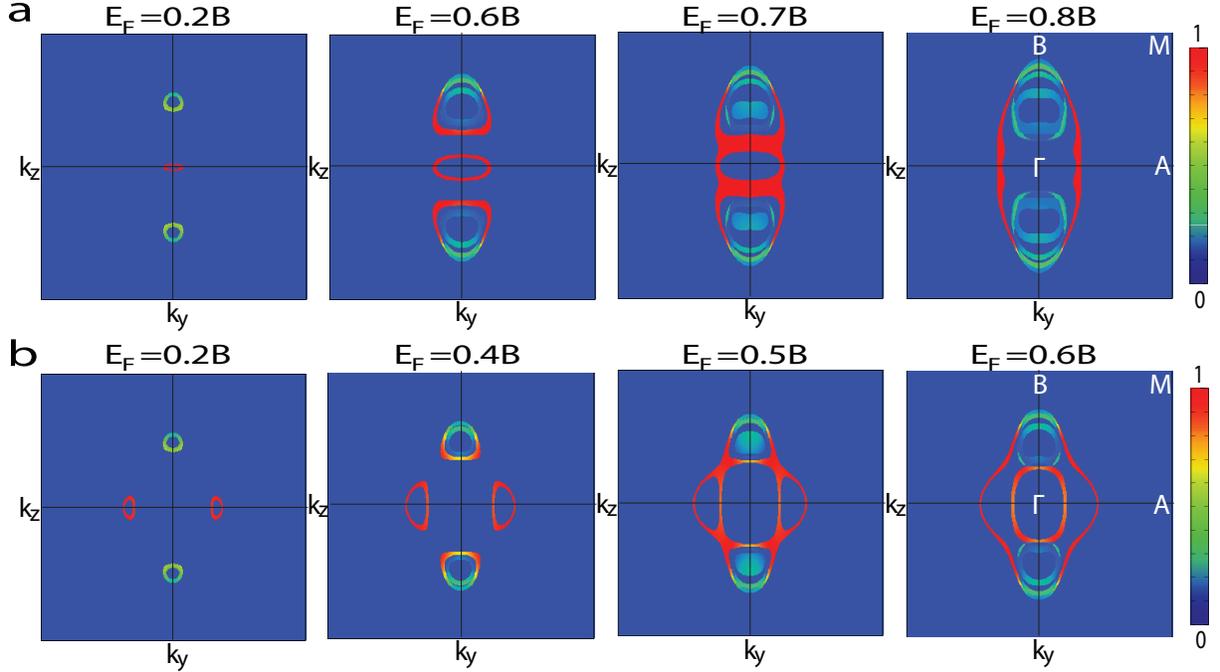}
\caption{
{\bf Evolution of the Fermi surface of a film whose surface normal direction is parallel to [100] direction
as a function of the Fermi energy.}
The wave function amplitudes confined within the first five layers from the top surface are plotted
for the states touching the Fermi level ($E_{F}$).
To obtain the Fermi surface we have solved numerically the lattice Hamiltonian in Equations (\ref{eqn:latticeH1})
and (\ref{eqn:latticeH2}).
({\bf a}) For a topological Dirac semimetal with $C_{4}$ symmetry, which has
$\nu_{2D}=1$ on the $k_{z}=0$ plane. The closed loop at the center of the surface BZ
is from a 2D surface Dirac point at the $\Gamma$ point. Two 3D bulk Dirac cones
also produce finite intensity symmetrically on the $k_{z}$ axis. As $E_{F}$
increases, the closed loop due to the 2D Dirac point deforms to a pair of Fermi arcs
connected to the bulk states.
({\bf b}) For a topological Dirac semimetal with $C_{4}$ symmetry, which has
$n_{m}=2$ on the $k_{z}=0$ plane. The two closed loops on the $k_{y}$ axis
are due to two 2D Dirac points localized on the surface. As $E_{F}$ increases, the closed loops due to
2D Dirac cones turn into four Fermi arcs.
In both ({\bf a}) and ({\bf b}), the 3D bulk Dirac fermions show the linear dispersion in the momentum space.
Hence, the number of surface Fermi arcs is solely determined by the 2D topological invariant on the
$k_{z}=0$ plane independent of the dispersion of the bulk Dirac fermions.
} \label{fig:surfacespectrum}
\end{figure*}


\end{document}